\documentclass{article}
\usepackage[utf8]{inputenc}
\usepackage{amsmath, bm}
\usepackage{graphicx}
\usepackage{changepage}
\usepackage{authblk}

\renewcommand{\thefootnote}{\fnsymbol{footnote}}
\title{Phase Separating Electrode Materials – Chemical Inductors?}

\author[1,2]{Klemen Zelič}
\author[1,2]{Igor Mele}
\author[3]{Arghya Bhowmik \thanks{arbh@dtu.dk}}
\author[1,2]{Tomaž Katrašnik \thanks{tomaz.katrasnik@fs.uni-lj.si}}
\affil[1]{University of Ljubljana, Faculty of Mechanical Engineering, SI-1000 Ljubljana, Slovenia}
\affil[2]{National Institute of Chemistry, Department of Materials Chemistry, 1000 Ljubljana, Slovenia}
\affil[3]{Department of Energy and Conversion Storage, Technical University of Denmark (DTU), Lyngby, 2800 Kgs, Denmark}

\date{} 

\begin{document}

\maketitle
\renewcommand{\thefootnote}{\arabic{footnote}}

We discover presence of chemical inductive effects in phase separating ion intercalation energy storage materials, specifically in lithium iron phosphate (LFP) and also lithium titanate oxide (LTO). These materials features fast (de)intercalation and slow diffusion relaxation phenomena which are prerequisites for observing such inductive effects. Presented finding is supported by the mechanistic model and analytical reasoning indicating that all equilibrium states that lay inside the miscibility gap of the phase separating material exhibit strong inductive response in the low frequency part of spectrum. We also explain why such inductive effects are not observed outside the miscibility gap. This letter presents the first mechanistic reasoning of previously reported electrode level experimental observation of inductance during impedance measurements at low currents.

\begin{center}
    \includegraphics[width=0.72\textwidth]{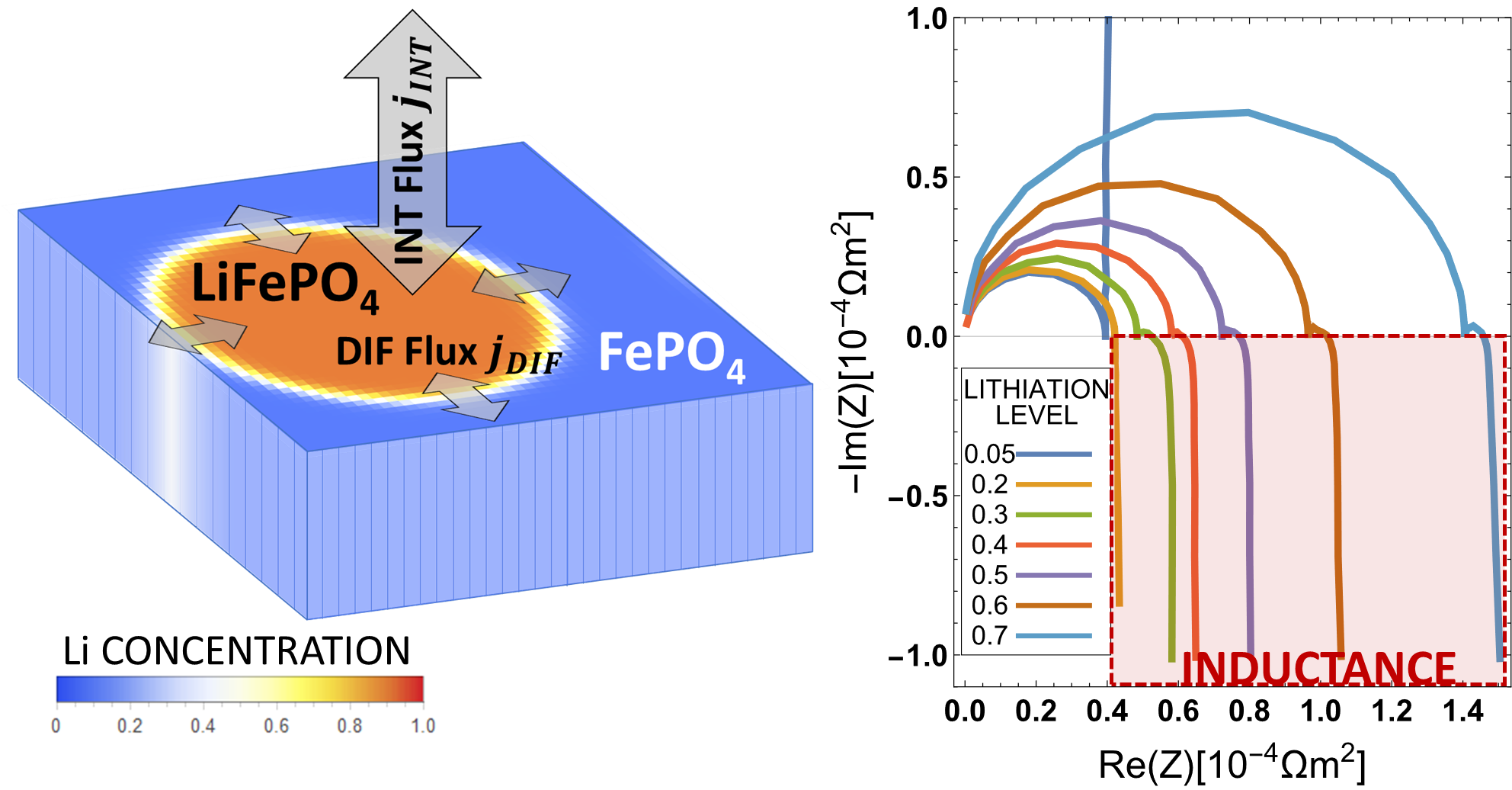}
\end{center}

Electrochemical impedance spectroscopy (EIS) is a powerful yet simple non-destructive experimental method that offers profound insight in the dynamic properties of studied system. It is widely used in the fields of chemistry, biology and material science \cite{chang2010electrochemical, orazem2008electrochemical, lasia2002electrochemical}. Accurate and precise physical model of the investigated system is crucial to utilize full potential of interpreting EIS spectra \cite{gabervsvcek2021understanding, ciucci2019modeling}. Without credible underlying models, interpretation of EIS spectrum can be questionable, especially when interpreting or discovering unconventional features in the spectrum.

Recently several inspiring articles reported a strong inductive response of different systems at very low frequencies (in the range of mHz), refereed to as "low frequency inductive loop" or "low frequency hook" \cite{klotz2019negative}. Reproducibility of these results for many different systems show that this is not a measuring artefact but rather an indication of general phenomena met in several different systems \cite{klotz2019negative}. Due to the lack of the underlying accurate physical models, interpretation of underlying processes remained challenging. 

Among several other systems, inductive effect was also measured in the low frequency part of spectrum also in lithium iron phosphate (LFP), intercalation cathode material for Li-ion batteries \cite{gaberscek2007meaning}. Due to the lack of plausible explanation of this measurement the published result was not given a significant attention. Inductive behaviour was later reported again on specific lithium - ion intercalation electrodes \cite{brandstatter2016myth, zhuang2009electrochemical}. However, article \cite{brandstatter2016myth} clearly reasons that presented inductive and negative capacitance loop in Swagelok type cells originates from springs, reference electrodes, drift, and corrosion, which are fundamentally different processes compared to inductive effects in ion
intercalation energy storage materials.

Klotz et al \cite{klotz2019negative} reviewed published experiments that observed inductive effects in low frequency part of EIS spectrum, summarizing two possible explanations that were prior given by \cite{bisquert2006negative, taibl2016impedance} and proposed a plausible empirical equivalent circuit model. Very recently it was proposed \cite{bisquert2022chemical} that chemical behaviour can also induce inductive effect if a system couples a fast conduction mode and a slowing down element. Based on this finding, authors of \cite{bisquert2022chemical} provide a generalised description of a generic system that exhibits chemical capacitance in a low frequency part of spectra. They coined the term "chemical inductor" and provided a basic mathematical formulation, requiring interaction of fast and slow phenomena, representing a necessary condition for a system to be a chemical inductor. Two articles following, by these authors, linked previously measured possible chemical inductive behaviours to the mathematical formulation presented in \cite{bisquert2022chemical} for halide perovskite memristors \cite{berruet2022physical}, FitzHugh–Nagumo neuron, the Koper–Sluyters electrocatalytic system, and potentiostatic oscillations of a semiconductor device \cite{bisquert2022hopf}. 

In this letter, we present a fundamental discovery on existence of chemical inductive effects in phase separating ion intercalation energy storage materials. In addition, we provide, for the first time, analytical reasoning and mechanistic model elucidating the entire causal chain from material specific properties of phase separating materials to its inductive effects. With this systematic analysis we reveal that phase separating energy storage materials exhibit significant inductive properties in the low frequency part of spectrum for all states within the miscibility gap. Specifically, inductive properties are demonstrated on the lithium iron phosphate (LFP) material, which is the only intercalation material where chemical induction was experimentally observed on the electrode level during impedance measurements at low currents \cite{gaberscek2007meaning}, whereas simulation results of lithium titanate Oxide (LTO) are also presented in Supplementary Information section S3 to demonstrate generality of the discovered phenomena. 

In addition to simulated results, we also provide analytical reasoning, why coupling slow phase boundary stabilization diffusion process, modelled by Cahn-Hilliard equation \cite{cahn1961spinodal,cahn1958free}, and fast destabilizing (de)intercalation, modelled by Butler-Volmer equation, e.g., \cite{newman2012electrochemical}, fulfill condition for a system to be a chemical inductor, when using parameters usually encountered in phase separating energy storage materials. Presented findings not only explain experimentally observed inductive behaviour of LFP in low frequency part of spectrum \cite{gaberscek2007meaning} but also answer a long standing fundamental question first asked by Srinivasan and Newman \cite{srinivasan2006existence}: "What is an impedance of phase separating electrode material inside miscibility gap?"

For the simulation of the spectra presented in this letter, a phase field model of phase separation (spinodal decomposition) \cite{cahn1961spinodal, cahn1958free} was chosen, that is widely applied and validated and accepted by society as accurate \cite{bai2011suppression, bazant2012phase, zelivc2019thermodynamically, fleck2018phase, wang2022reaction}. Both materials were simulated by a widely utilized coupled Cahn-Hilliard equations \cite{cahn1961spinodal,cahn1958free}, to model intra-particle diffusion, and Butler-Volmer equation, e.g., \cite{newman2012electrochemical}, to model (de)intercalation process solved in two dimensions. Results clearly reveal that only intermediate lithiation levels - inside miscibility gap - exhibit chemical induction characteristics. This is analytically and numerically supported by the fact that two competing processes with significantly different characteristic times are present in the system - fast destabilizing (de)intercalation and slow phase boundary stabilization diffusion process.

Key governing equations of the model are summarized below, while detailed description is given in Supplementary material section S1. Spatial and temporal intra-particle fields of Li concentration and chemical potential are modelled with an established phase field approach \cite{bai2011suppression, bazant2012phase, zelivc2019thermodynamically, fleck2018phase, wang2022reaction} consisting of coupled equations \ref{eqn:Cahn-Hilliard} to \ref{eqn:j_tot}. This is realized by coupling the Cahn-Hilliard equation (eq. \ref{eqn:Cahn-Hilliard}) \cite{cahn1961spinodal}, equation for modelling chemical potential (eq. \ref{eqn:chemical potential}) \cite{cahn1958free} incorporating terms for regular solution entropy and mixing enthalpy, phase field gradient penalty and strain energy, and appropriate boundary conditions (equations \ref{eqn:Butler-Volmer}, \ref{eqn:BCx} and \ref{eqn:BCy})

\begin{equation}
\label{eqn:Cahn-Hilliard}
\frac{\partial c(\mathbf{r},t)}{\partial t} = \nabla \frac{ c(\mathbf{r},t)}{RT} \mathbf{D} \nabla \mu(\mathbf{r},t) + \frac{1}{d_p}j_{INT},
\end{equation}

\begin{equation}
\label{eqn:chemical potential}
\mu(\mathbf{r},t) = RT \ln{\frac{c(\mathbf{r},t)}{c_m-c(\mathbf{r},t)}} + \Omega \left(1 - \frac{2c(\mathbf{r},t)}{c_m}\right) - \nabla \bm{\kappa} \nabla c(\mathbf{r},t) + B_0\left[\frac{c(\mathbf{r},t)-\Bar{c}(t)}{c_m^2}\right].
\end{equation}

System of equations \ref{eqn:Cahn-Hilliard} and \ref{eqn:chemical potential} was solved on the two dimensional domain, since diffusion in LFP is much faster along (010) crystallographic direction compared to other directions. This approach is consistent with published articles \cite{bai2011suppression, zelivc2019thermodynamically}, while selection of model dimensionality does not limit generality of reported findings. 

Dimensionality of the model, also influences formulation of boundary conditions. Consequently, (de)intercalation process of lithium across the particle faces perpendicular to (010) crystallographic direction, i.e., source term $\frac{1}{d_p}j_{INT}$ in equation \ref{eqn:Cahn-Hilliard}, was described using Butler-Volmer flux, driven by potential difference ($\phi$) between the LFP material and the surrounding electrolyte
\begin{equation}
\label{eqn:Butler-Volmer}
j_{INT} = j_0\left[\exp\left(\frac{F\alpha}{RT}\left(\phi(t)-\frac{\mu(\mathbf{r},t)}{F}\right)\right) - \exp\left(\frac{F(1-\alpha)}{RT}\left(\phi(t)-\frac{\mu(\mathbf{r},t)}{F}\right)\right)\right].
\end{equation}
In the directions perpendicular to the (010) crystalographic direction, periodic boundary conditions were used.
\begin{equation}
\label{eqn:BCx}
c(x=0)\nabla\mu(x=0) = c(x=L)\nabla\mu(x=L),
\end{equation}

\begin{equation}
\label{eqn:BCy}
c(y=0)\nabla\mu(y=0) = c(y=L)\nabla\mu(y=L).
\end{equation}

\begin{figure}[h!]
    \centering
    \includegraphics[width=\textwidth]{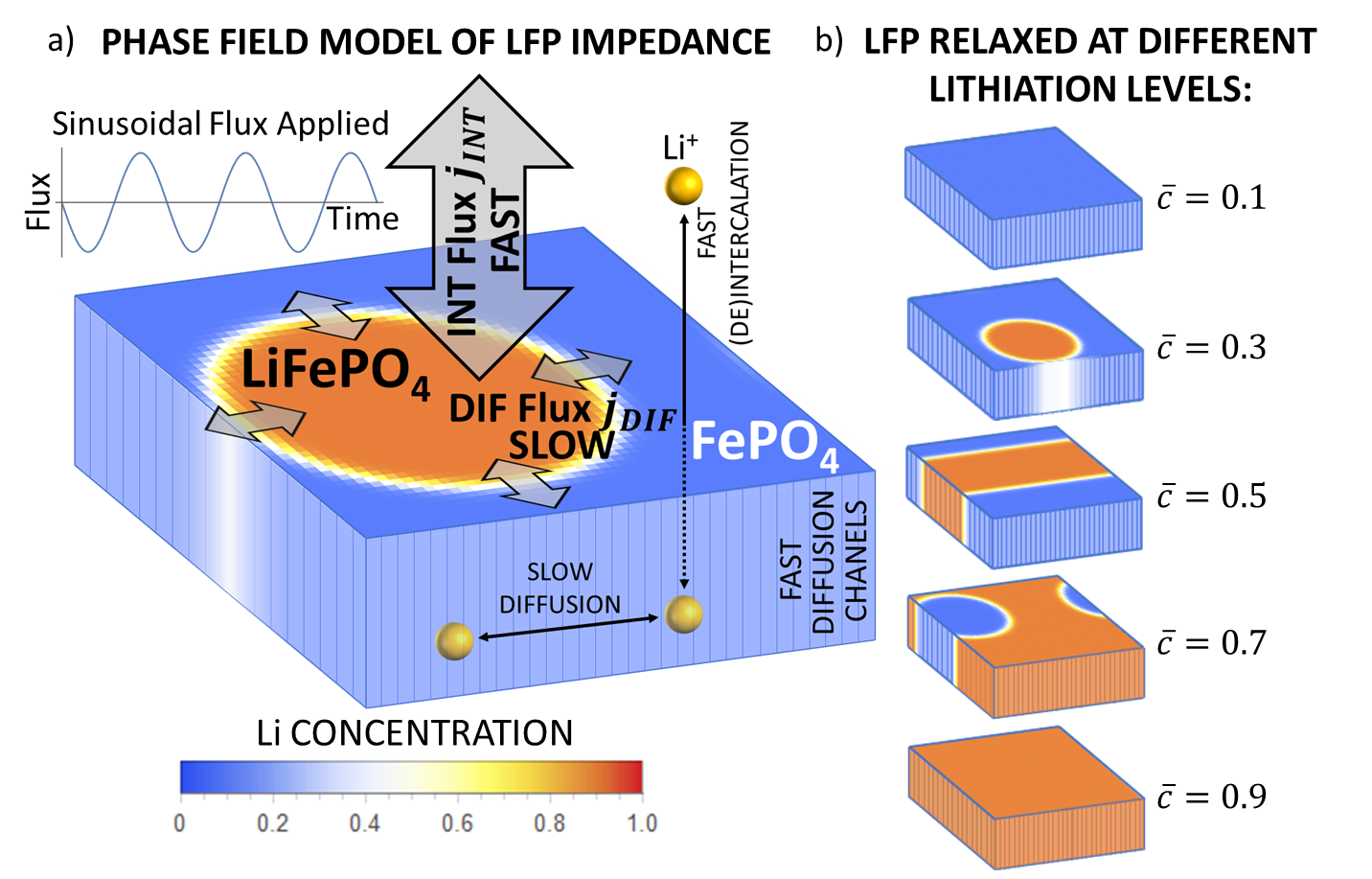}
    \caption{a) 
     Schematic representation of the model used for simulating impedance spectra. (De)Intercalation flux $j_{INT}$ is modeled by Butler-Volmer equation, perpendicular to the domain whereas uphill diffusion and phase separation flux $j_{DIF}$ in the domain plane is modeled by Cahn-Hilliard equation. The magnitude of $j_{INT}$ is large in comparison to $j_{DIF}$, representing two processes on very different time scales: fast destabilizing (de)intercalation and slow diffusion relaxation that stabilizes phase boundaries. Relaxed structure of phase separated LFP used as initial condition for the impedance simulation is represented as a color map, where red tones represent high concentration of lithium (Li-rich phase $LiFePO_4$), blue tones represent low concentration of lithium (Li-poor phase $FePO_4$) and white and yellow tones represent intermediate concentrations at phase boundaries. b) Relaxed LFP states for five different lithiation levels. Presented structures were obtained by relaxation of slightly perturbed constant concentration field at zero flux. For highest and lowest lithiation levels (0.95 and 0.05) initial condition differs drastically from all other initial conditions. These two values of lithiation namely lay outside of the LFP miscibility gap. For these two cases homogeneous distribution of lithium concentration represent an equilibrium state, wheres for all other initial condition equilibrium states exhibit a phase separation to Li-rich and Li-poor phases.}
    \label{fig:model}
\end{figure}

To derive a credible model for modeling EIS, double layer effects need to be considered in addition to (de)intercalation processes. Therefore, in addition to (de)intercalation flux (source term $\frac{1}{d_p}j_{INT}$ in equation \ref{eqn:Cahn-Hilliard}) also contribution of a flux into (out of) the double layer is considered. This is realized via the equation proposed and elaborated in references \cite{newman1975porous, meyers2000impedance}
\begin{equation}
\label{eqn:Double-Layer}
j_{DL} = \frac{C_{DL}}{F}\frac{\partial \phi(t)}{\partial t}
\end{equation}
where $C_{DL}$ denotes the double layer capacitance per surface area. 

Consequently, a total flux of Li and thus also electrons equals
\begin{equation}
\label{eqn:j_tot}
j_{TOT} = j_{INT}+j_{DL}.
\end{equation}

Simulations were performed, consistent with experimental EIS procedure, by applying periodic sinusoidal flux ($j_{TOT}$) with low amplitude to the relaxed LFP structures and solved on two dimensional domain as presented in Figure \ref{fig:model}.  

Impedance of the system was calculated as the ratio of potential difference response of the system ($\phi$) and imposed electric current $I$ corresponding to applied flux $j_{TOT}$ as 
\begin{equation}
\label{eqn:Impedance}
Z = \frac{\phi}{F\int_A j_{TOT} dA }.
\end{equation}
All the symbols used in equations are listed in Table \ref{tab:table}.

\begin{table}
    \centering
    \caption{Significance of symbols used in equations \ref{eqn:Cahn-Hilliard} to \ref{eqn:Impedance}}
    \label{tab:table}

\begin{center}
\begin{tabular}{ l c l }
 \bf{Symbol} &  \bf{Units}  &  \bf{Physical Significance} \\
 \hline
 \hline
$A$ &  m$^2$& computational domain area \\
$\alpha$ &  - & charge transfer coefficient\\
$B_0$ &  Pa & strain coefficient\\
$c$ & mol/m$^3$ & molar concentration of lithium \\  
$\bar{c}$ & mol/m$^3$ & average molar concentration of lithium \\
$C_{DL}$ & F/m$^2$ & double layer specific capacitance \\
$c_m$ & mol/m$^3$ & maximal molar concentration of lithium in LFP\\  
$\mathbf{D}$ & m$^2$/s & diffusivity tensor\\ 
$d_p$ & m & domain thickness\\
$F$ & As/mol & Faraday constant\\
$\phi$ & V & potential difference between LFP and electrolyte\\
$j_0$ & mol/m$^2$s & Butler-Volmer exchange flux density\\
$j_{DIF}$ & mol/m$^2$s & diffusion flux density\\
$j_{INT}$ & mol/m$^2$s & (de)intercalation flux density\\
$j_{DL}$ & mol/m$^2$s & double layer flux density\\
$\kappa$ & J/m & gradient penalty parameter\\
$L$ & m & domain size\\
$\mu$ & J/mol & chemical potential\\    
$\mathbf{r} = (x,y)$ & m & position vector\\ 
$R$ & J/mol K & gas constant\\
$S$ & mol/m$^3$s & source term\\
$t$ & s & time\\
$T$ & K & temperature\\ 
$Z$ & $\Omega$ & impedance\\
$\Omega$ & J/mol & regular solution parameter\\

\end{tabular}
\end{center}
\end{table}

Impedance spectra of the LFP active material were simulated in the frequency interval from $1$ $mHz$ to $100$ $kHz$, for nine different lithiation levels of LFP material. These lithiation levels were selected in a way to yield concentrations inside and outside of the miscibility gap.

\begin{figure}[h!]
\hspace*{-3cm}    
    \centering
    \includegraphics[width=1.42\textwidth]{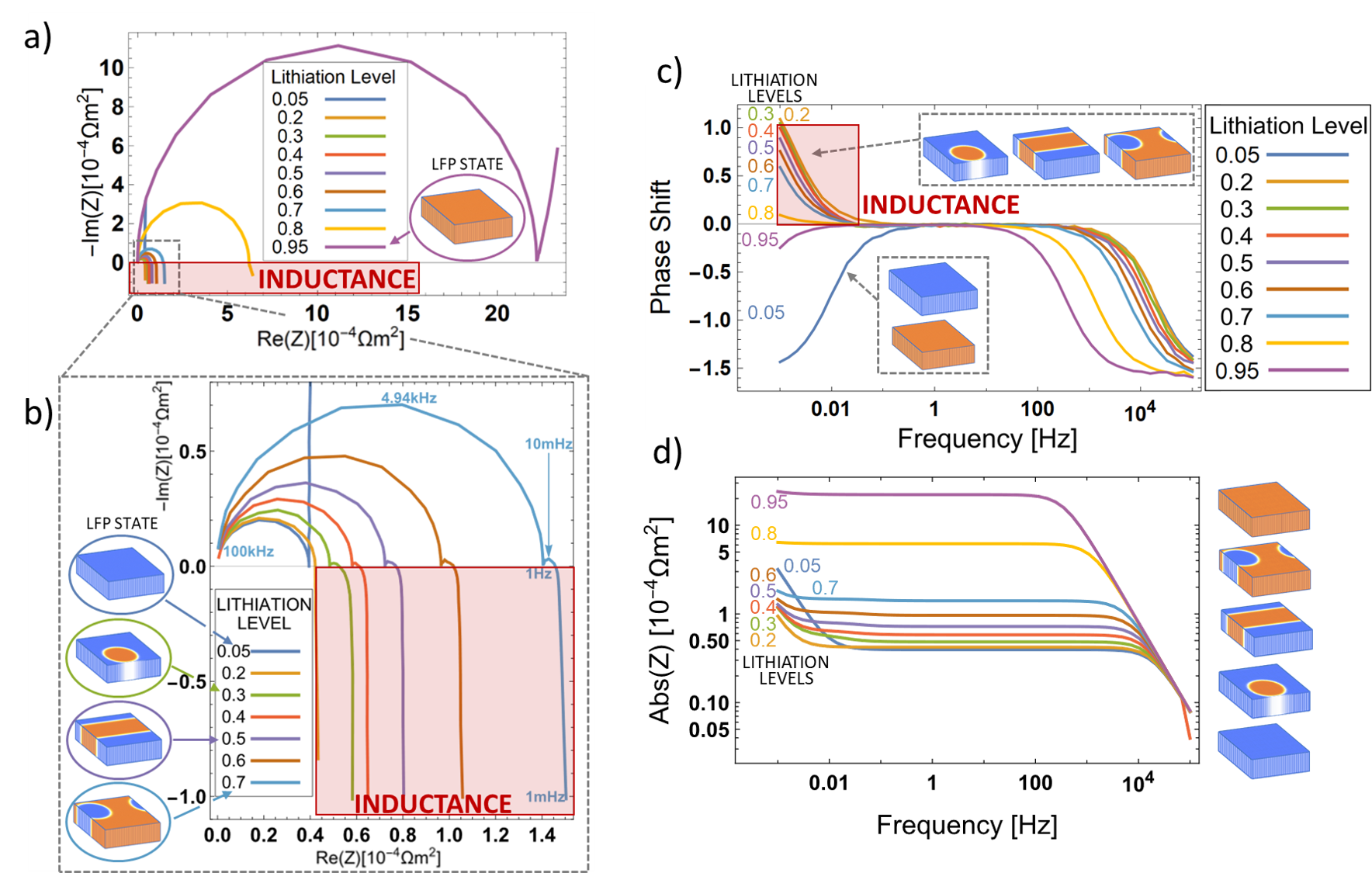}
    \caption{a) Simulated Nyquist plots of the LFP material at nine different lithiation levels. Inductive effects in low frequency part of spectrum are seen for lithiation levels within the miscibility gap (i.e. lithiation levels from 0.2 to 0.8), while lithiation levels outside the miscibility gap (i.e. lithiation levels 0.05 and 0.95) exhibit capacitive and not inductive effects. b) Zoom in of the image a) to the lithiation levels from 0.05 to 0.7. Chemical inductance is well seen in all curves apart from one corresponding to lowest lithiation level outside the miscibility gap. Images on the left show the relaxed LFP states that were used as initial conditions for the simulation of impedance (Figure 1b)). Phase separated initial conditions result in chemical inductive effects, whereas non-phase separated initial condition cases exhibit only capacitive effects. c) Bode diagram of phase shift dependence on frequency, corresponding to the Nyquist plot curves from image a) clearly showing the difference between non-phase separated and phase separated initial LFP states in the low frequency part of the spectrum. d) Absolute value of impedance as a function of frequency for nine different lithiation levels.}
    \label{fig:impedance}
\end{figure}

Results of EIS are presented in Figure \ref{fig:impedance}. Figure clearly reveals expected capacitive arcs in the high frequency part of the spectra. These capacitive arcs are a consequence of the double layer response. At the low frequency part of spectra, strong inductive effects are seen in simulation result for all lithiation levels inside the miscibility gap, while the two lithiation levels outside the miscibility gap, i.e., 0.05 and 0.95, do not feature inductive effects.  

According to \cite{bisquert2022chemical} the necessary condition for the system to exhibit chemical induction are two competing processes: fast destabilizing and slow stabilizing process. In the case of phase separation in intercalation electrode, these two processes correspond to fast destabilizing (de)intercalation and slow phase boundary stabilization diffusion process, where phase boundary is established due to uphill diffusion, which is in the present case modelled with the Cahn-Hilliard equation \ref{eqn:Cahn-Hilliard}. The influence of both described processes on the system impedance can be seen from expression defining total current in to the domain, obtained by reformulating the total flux equation \ref{eqn:j_tot} and written in the limit of small harmonic perturbation as: 
\begin{equation}
\label{eqn:Current}
I = F\int_A\hat{j}_{TOT}dA = AC_{DL}\frac{\partial \hat{\phi}}{\partial t} + \frac{F^2Aj_0}{RT}\hat{\phi} + \frac{Fj_0}{RT}\int_A \hat{\mu}dA,
\end{equation}
where $\hat{ }$ symbol over the variable denotes the small perturbation (detailed derivation is provided in Supplementary material S2). First term on the right-hand side of the equation \ref{eqn:Current} arises from double layer equation \ref{eqn:Double-Layer} and the second and the third terms are linearised and integrated (de)intercalation flux from equation \ref{eqn:Butler-Volmer}. These two terms describe the fast, destabilising process in the system with a characteristic time of few milliseconds. Slow stabilizing process with the characteristic time of few seconds is described with the last term on the right hand side of the equation \ref{eqn:Current}, defined with integral of chemical potential which adapts with the long characteristic time of phase boundary stabilization diffusion process. The time dependence of this term and corresponding slow dynamics can be seen from reformulated form of equation \ref{eqn:Cahn-Hilliard}, which provides the integro-differential equation defining the chemical potential surface integral from equation \ref{eqn:Current}. Integration of the equation \ref{eqn:Cahn-Hilliard} together with Leibniz integral rule and taking into account that $\hat{c} = \frac{\partial \hat{c}}{\partial \hat{\mu}}\hat{\mu}$ and $\nabla \hat{c} = \frac{\partial \hat{c}}{\partial \hat{\mu}}\hat{\mu}$ gives
\begin{equation}
\label{eqn:Cahn-Hilliard_mu}
\frac{\partial }{\partial t}\int_A \hat{\mu}dA = \int_A \left[ \frac{D}{RT}\hat{\mu}\nabla^2\hat{\mu} + \frac{D}{RT}(\nabla\hat{\mu})^2 + \frac{1}{d_p}\hat{j}_{INT}\right]dA. 
\end{equation}

Equation \ref{eqn:Cahn-Hilliard_mu} provides key insight into observing inductive effects due to the interaction between both competing processes. Terms with the chemical potential gradient ($\nabla\hat{\mu}$) describe phase boundary stabilization diffusion and the source term $\frac{1}{d_p}\hat{j}_{INT}$ describes (de)intercalation. When LFP material is relaxed inside miscibility gap (presented cases for lithiation levels from 0.2 to 0.8 in Figure \ref{fig:model} b)) it phase separates. Application of external flux to the phase separated LFP state, result in (de)intercalation source term $\frac{1}{d_p}\hat{j}_{INT}$ that is not constant across the whole domain but rather governed by in-homogeneous concentration field. This produces chemical potential gradients in the system and slow diffusion of lithium across the domain that stabilizes phase boundaries follows as a direct consequence of fast destabilizing (de)intercalation process. These processes explain the occurrence of inductive effect in the low part of spectrum for LFP inside miscibility gap. This observation can also be confirmed by the fact that system of equations \ref{eqn:Current} and \ref{eqn:Cahn-Hilliard_mu} satisfy mathematical condition for chemical inductor postulated in \cite{bisquert2022chemical} for lithiation levels within the miscibility gap. 

\begin{figure}[h!]
    \centering
    \includegraphics[width=\textwidth]{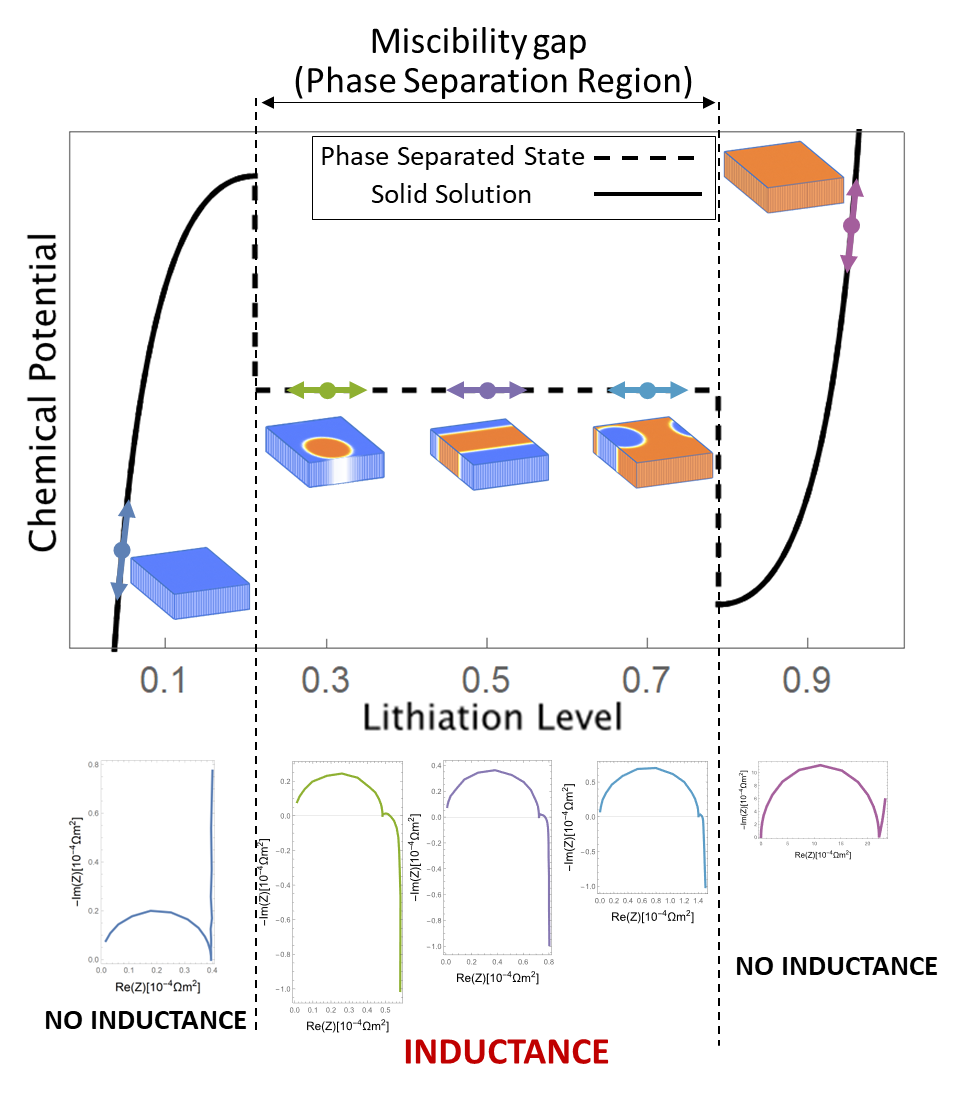}
    \caption{a) Single particle chemical potential dependence on lithiation level for LFP material, for the case of solid solution (solid line) and phase separated state (dashed line). The representation of the solid solution chemical potential outside the miscibility gap and chemical potential of the phase separated state inside the miscibility gap is consistent with references \cite{zelivc2019thermodynamically, katrasnik2021entering, malik2013critical, li2014current}. Dots on the plot show the lithiation levels of different equilibrium LFP states and arrows around the dots show schematically the harmonic oscillation of lithiation level around the equilibrium point during the EIS experiment. Particles with the lithiation levels inside the miscibility gap exhibit inductance effects in the low frequency part of spectra.}
    \label{fig:spinodal}
\end{figure}

Outside miscibility gap (cases for lithiation level 0.05 and 0.95 in Figure \ref{fig:model} b)) chemical potential and concentration fields are homogeneous in the relaxed state, since the system is in the solid solution state (Figure \ref{fig:spinodal}). Consequently, applied (de)intercalation flux is (nearly fully) homogeneous, preserving gradient-free chemical potential and, hence, yielding no redistribution of lithium in the domain. Therefore, no stabilising slow diffusion process accompany fast (de)intercalation in this regime. Consequently, the two terms with chemical potential gradient ($\nabla\hat{\mu}$) in equation \ref{eqn:Cahn-Hilliard_mu} limit towards zero. System of equations \ref{eqn:Current} and \ref{eqn:Cahn-Hilliard_mu} transforms in such a manner that they do not satisfy the condition for the chemical inductor \cite{bisquert2022chemical}, which is also fully consistent with simulated results \ref{fig:impedance}.  

Presented results and analytic derivations clearly indicate that phase separating nature of the LFP is of the crucial importance for its behaviour as the chemical inductor. The reason for chemical induction to exist in such a system namely arises from a significant difference between the characteristic times for fast lithium (de)intercalation and slow in-plane diffusion that stabilizes phase boundaries. In a similar system without phase separation, no slow stabilizing diffusion process would take place during EIS analysis, which removes the long relaxation time scale of the system, as presented in Figure \ref{fig:impedance}. Figure \ref{fig:impedance} thus clearly shows that for the case of lithiation levels outside miscibility gap (lithiation levels 0.05 and 0.95) the system exhibits capacitive nature at low frequencies instead of inductive nature due to absence of phase separation.

Similar inductive behaviour, as disclosed for the LFP material, can be anticipated also in other electrode materials with phase separation that are characterized by fast destabilizing (de)intercalation and slow phase boundary stabilizing diffusion relaxation (e.g. LiMnPO$_4$ \cite{aravindan2013limnpo}, NaFePO$_4$ \cite{oh2012reversible}, Li$_4$Ti$_5$O$_12$ (LTO) \cite{sun2018electrochemical}). In the Supplementary Information section S3 inductor behaviour of such materials and thus generality of the observed phenomena is demonstrated also for the LTO, which confirms that phase separation is a key phenomenon to observe inductor behaviour. Electrode materials that undergo structural phase transitions during (de)lithiation (e.g. NMC and NCA \cite{mohanty2016modification}) do namely not feature inductor effects. Crystal structure phase transitions, which do not result in a separation to Li-poor and Li-rich domains, do not represent a necessary condition for material to be chemical inductor as discussed above, as slow diffusion relaxation associated with movement of the phase boundary is not present. 
 
This letter for the first time associates and reasons interrelation between necessary condition to exhibit inductive phenomena \cite{bisquert2022chemical} and real processes in energy storage materials establishing a causal interrelation between material specific properties and inductive effects. Thereby, it solves long lasting challenge about the impedance of phase separating materials inside miscibility gap \cite{srinivasan2006existence}. Results presented in this letter show that impedance response of a single phase separating particle exhibits inductive effect inside miscibility gap which is the first mechanistic reasoning of previously reported electrode level experimental observation \cite{gaberscek2007meaning} during impedance measurements at low currents.

\bibliography{bibliography} 
\bibliographystyle{unsrt}

\end{document}


\maketitle
\renewcommand{\thefootnote}{\arabic{footnote}}
\renewcommand{\thesection}{S\arabic{section}}
\renewcommand{\theequation}{S\arabic{equation}}
\renewcommand{\thefigure}{S\arabic{figure}}

\section{Model}
\subsection{Governing equations}
For the simulation of LFP impedance spectra widespread phase field modeling approach was used, that describes chemical potential and concentration field inside LFP material, which govern the impedance. Standard phase field approach was followed, where gradient penalty term is summed to the free energy density of the system to obtain the functional defining total free energy \cite{cahn1961spinodal, cahn1958free, bazant2012phase, wang2020application, t2016phase}.
\begin{equation}
\label{eqn:Total_F}
    \mathcal{F}=\int_V\left[f(c(\mathbf{r},t)) + \frac{\kappa}{c_m^2}[\nabla c(\mathbf{r},t)]^2\right]dV
\end{equation}
$\mathcal{F}$ denotes total free energy of the system, $V$ is domain volume, $c$ is concentration (phase field parameter) that depends on position $\mathbf{r}$ and time $t$. $\kappa$ represents gradient penalty coefficient and $c_m$ is maximal concentration of lithium in the system. Function $f$ is free energy density of the system, that can be written as
\begin{equation}
\label{eqn:f}
    f(c) = RTc\ln{\left(\frac{c}{c_m}\right)} + RT\left(c_m-c\right) \ln{\left(\frac{c_m-c}{c_m}\right)} + \Omega\frac{c(c_m-c)}{c_m^2} + \frac{B_0}{2c_m^2}(c-\Bar{c})^2.
\end{equation}
$R$ is gas constant, $T$ temperature, $\Omega$ regular solution parameter, $B_0$ is elastic strain energy of interface and $\Bar{c}$ is lithiation level of the system. First three terms in the free energy density are obtained from regular solution theory \cite{cahn1961spinodal, cahn1958free, zelivc2019thermodynamically}. The use of regular solution to describe the phase separating materials that undergo spinodal decomposition (including LFP) is established and well validated approach \cite{wang2020application}. Last term describe a strain energy in the system. Appropriately for the simulation of EIS spectrum (small harmonic perturbation around equilibrium),  description of free energy contribution due to the deformation is obtained by evaluation of strain energy for small fluctuation of the system around equilibrium \cite{cahn1958free, cogswell2012coherency}:
\begin{equation}
\label{eqn:strain}
   \frac{1}{2}C_{ijkl}\varepsilon_{ij}\varepsilon_{kl}\approx\frac{B_0}{2c_m^2}(c-\Bar{c})^2,
\end{equation}
where $C_{ijkl}$ represent stiffness tensor and $\varepsilon_{ij}$ is strain field. The same approach to the description of strain in phase separating materials including LFP was used in several other studies \cite{cahn1961spinodal, zelivc2019thermodynamically, cogswell2012coherency, fleck2018phase}.

Minimisation of total free energy from equation \ref{eqn:Total_F} with Lagrange variation principle result in the chemical potential written in Equation 2 in main paper. Together with the first Fick's law of diffusion 
\begin{equation}
\label{eqn:first_fick}
   \mathbf{j} = -\frac{\mathbf{D} c}{RT}\nabla \mu
\end{equation}
and continuity equation
\begin{equation}
\label{eqn:Continuity}
   \frac{\partial c}{\partial t} = -\nabla \cdot \mathbf{j}
\end{equation}
Cahn-Hilliard system of the equations is obtained that govern the dynamic properties of the modeled system. $\mathbf{j}$ represent molar flux of lithium in the domain and $\mathbf{D}$ is diffusivity tensor. Taking in to account the fact that LFP diffusivity tensor is highly anisotropic (diffusion in (010) crystallographic direction is orders of magnitude faster than any other direction \cite{zhang2011structure}), equation \ref{eqn:Continuity} can be written as 
\begin{equation}
\label{eqn:Continuity2D}
   \frac{\partial c}{\partial t} = -\nabla \cdot \mathbf{j}_{DIF} + \frac{j_{INT}}{d_p}
\end{equation}
where $\mathbf{j}_{DIF}$ now denotes the diffusion fluxes in two directions perpendicular to the (010) direction (denoted later as $x$ and $y$ direction), $\nabla$ two dimensional divergence operator in directions $x$ and $y$, $d_p$ is particle thickness in to the direction parallel to (010) and $j_{INT}$ is (de)intercalation flux across particle interface perpendicular to (010). (De)Intercalation flux was reshaped in to the form of source term in equation \ref{eqn:Continuity2D}. Inserting equation \ref{eqn:first_fick} in to the \ref{eqn:Continuity2D} gives final dynamic equation of the system (equation 1 in the main paper).

Interfacial flux $j_{INT}$ in the source term that represent an electrochemical reaction at the surface of the particle was modeled by Butler-Volmer equation (eq. 3 in the main paper). Periodic boundary conditions were used in the directions perpendicular to (010) - equations 4 and 5 in main paper. 

Additionally small fluctuations in chemical potential were introduced in to the model corresponding to the thermal fluctuations. This was done the same way as presented in paper \cite{bai2011suppression}.

\subsection{Equation Solving}
Governing equations of the model were solved by finite volume method on the two dimensional domain in the plane perpendicular to (010) crystalographic direction, discretized to 50x50 mesh. IDA solver from open source Sundials solver package was used to solve obtained ODE system of equations. Chemical potential and concentration in every control volume of the mesh were obtained as a result.

\subsection{Validation}
In this chapter few key results are gathered that show validity and plausibility of used model, by comparison of the simulated results to the prior art. 

\begin{figure}
    \centering
    \includegraphics[width=\textwidth]{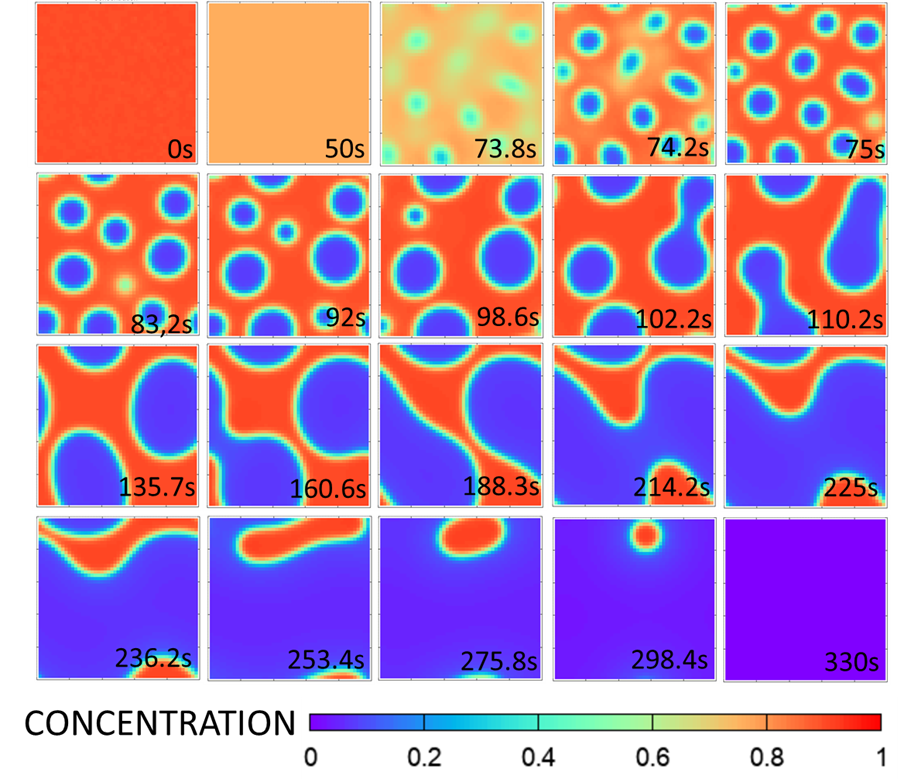}
    \caption{Time dependency of concentration field inside the domain during constant flux delithiation that corresponds to 10C charge. Time stamps of each frame are in the bottom right corners. Red tones indicate high concentration of lithium (Li-rich phase) and blue tones indicate low concentration of lithium (Li-poor phase)}
    \label{fig:delithiation}
\end{figure}

Figure \ref{fig:delithiation} shows a sequence of 2D concentration plots in case of active material delithiation across the whole depth of discharge (DoD) range. 
The transition from initial solid solution to the phase separated state during delithiation is seen in \ref{fig:delithiation}. Transition from solid solution to phase separated regime is well seen (frames from 50s to 75s) compatible with the results presented in references i.e. \cite{zelivc2019thermodynamically, bai2011suppression}. Phase separation is followed by domino cascade progress of concentration field evolution \cite{delmas2011lithium}. Similar time traces were reported for the case of LFP by several authors i.e. \cite{zelivc2019thermodynamically, bai2011suppression, bazant2013theory, singh2008intercalation, kao2010overpotential, burch2008phase, delmas2011lithium, cogswell2012coherency, zeng2013cahn, tang2009model, tang2010electrochemically, di2014cahn, dargaville2013persistence}.

\begin{figure}[h]
    \centering
    \includegraphics[width=\textwidth]{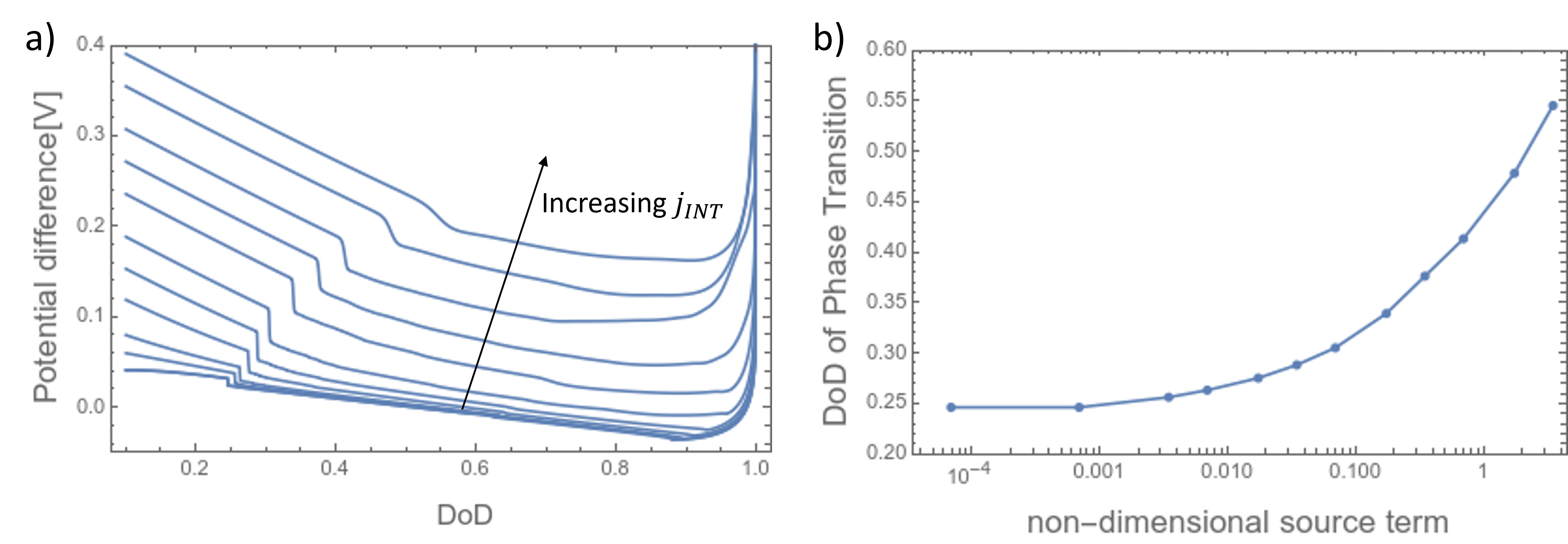}
    \caption{a) DoD dependency of potential difference $\phi$ for eleven different values of deintercalation flux $j_{INT}$. b) DoD values at phase transition dependency on deintercalation flux $j_{INT}$.}
    \label{fig:suplementary2}
\end{figure}

Further, the magnitude of constant intercalation flux was varied in order to test the model. Potential difference $\phi$ for different magnitudes of intercalation flux are plotted in Figure \ref{fig:suplementary2} a). The DoD value of phase separation is well seen in the plots of potential difference in Figure as \ref{fig:suplementary2} a) steps. The DoD values of phase transitions were determined in post processing and plotted versus deintercalation flux in Figure \ref{fig:suplementary2} b). 
Both results from Figure \ref{fig:suplementary2} coincide well with previously published results in references \cite{bai2011suppression, ichitsubo2013elastically, zeng2014phase}.

\section{Linearisation}
This section serves as a supplementary to the equation 9 and 10 of main paper, where linearised form of different quantities is used denoted in the main paper with the symbol $\hat{}$ over the variables. The linearisation is defined as transforming the functional dependency to its first term in the corresponding Taylor series expansion:
\begin{equation}
g(u)\approx\frac{\partial g(u_0)}{\partial u}g(u-u_0),
\end{equation}
where $g$ is arbitrary function of independent variable $u$ and $u_0$ is an arbitrary value of a variable $u$.

(De)Intercalation flux $j_{INT}$ modeled by Butler-Volmer equation was linearized for low perturbation in potential difference $\hat{\phi}$ as:
\begin{multline}
\hat{j}_{INT} = j_0\left[\exp\left(\frac{F\alpha}{RT}\left(\hat{\phi}-\frac{\hat{\mu}}{F}\right)\right) - \exp\left(\frac{F(1-\alpha)}{RT}\left(\hat{\phi}-\frac{\hat{\mu}}{F}\right)\right)\right]\approx\\
\approx \frac{Fj_0}{RT}\left(\hat{\phi} - \frac{\hat{\mu}}{F}\right)
\end{multline}

\section{LTO}
\begin{figure}[h!]
    \centering
    \includegraphics[width=0.5\textwidth]{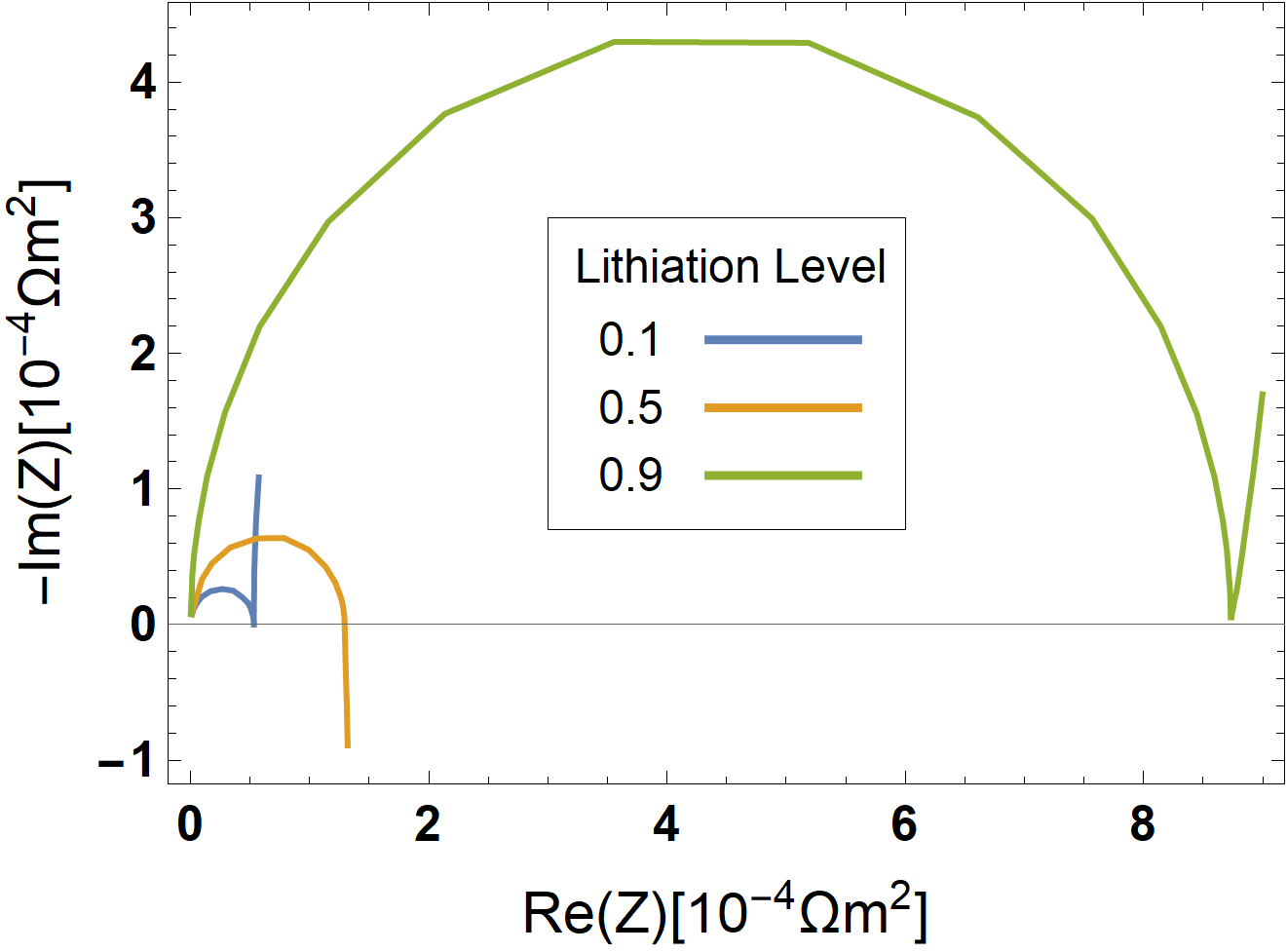}
    \caption{Niquist plot for impedance spectrum of LTO single particle for three different lithiation levels. Parameters for LTO were taken from reference \cite{vasileiadis2018toward}}
    \label{fig:LTO}
\end{figure}

Figure \ref{fig:LTO} shows the chemical inductance in LTO electrode material. Model presented in main paper and section S1 was used for simulations on LTO, parameterized with the values given in reference \cite{vasileiadis2018toward}. The same trend of chemical inductance inside miscibility gap and capacitance out of the miscibility gap can be seen from the figure.

\bibliographystyle{unsrt}
\bibliography{refs}